\documentclass[twoside,twocolumn,english,aps,prb,reprint]{revtex4-2}
\usepackage[T1]{fontenc}
\usepackage[latin9]{inputenc}
\usepackage{color}
\usepackage{babel}
\usepackage{amsmath}
\usepackage{amsthm}
\usepackage{amssymb}
\usepackage{graphicx}
\usepackage[unicode=true, allcolors=blue,pdfusetitle,
 bookmarks=true,bookmarksnumbered=false,bookmarksopen=false,
 breaklinks=false,pdfborder={0 0 1},backref=false]{hyperref}

\makeatletter

\usepackage{easybmat}
\usepackage{bm}
\usepackage{graphicx}
\def\sqz{.5}
\def\sqzinverse{2}
\def\hsqz{.7}

\newlength\bls
\newlength\tmplength
\def\scalebrace#1#2{\tmplength=#1\bls\relax%
  \scalebox{\hsqz}[\sqz]{\rotatebox{90}{$\underbrace{\hspace{\sqzinverse\tmplength}}$}}%
  \raisebox{\dimexpr+.5\tmplength+.5\dp\strutbox-.5\ht\strutbox}{$\scriptstyle \; #2$}}

\makeatletter
\renewcommand*\env@matrix[1][*\c@MaxMatrixCols c]{%
  \hskip -\arraycolsep
  \let\@ifnextchar\new@ifnextchar
  \array{#1}}

\makeatother

\renewcommand\[{\begin{equation}}
\renewcommand\]{\end{equation}}

\usepackage{times}

\makeatother

\begin{document}
\title{Phase-transition-like behavior in information retrieval of a quantum scrambled random circuit system}
\author{J.-Z. Zhuang}
\email{zhuangjz21@mails.tsinghua.edu.cn}
\affiliation{Center for Quantum Information, Institute for Interdisciplinary Information Sciences, Tsinghua University, Beijing 100084, PR China}

\author{Y.-K. Wu}
\affiliation{Center for Quantum Information, Institute for Interdisciplinary Information Sciences, Tsinghua University, Beijing 100084, PR China}

\author{L.-M. Duan}
\email{lmduan@tsinghua.edu.cn}
\affiliation{Center for Quantum Information, Institute for Interdisciplinary Information Sciences, Tsinghua University, Beijing 100084, PR China}

\begin{abstract}
Information in a chaotic quantum system will scramble across the system,
preventing any local measurement from reconstructing it. The scrambling
dynamics is key to understanding a wide range of quantum many-body systems.
Here we use Holevo information to quantify the scrambling dynamics,
which shows a phase-transition-like behavior. When applying long random Clifford circuits
to a large system, no information can be recovered from a subsystem
of less than half the system size. When exceeding half the system
size, the amount of stored information grows by two bits of classical
information per qubit until saturation through another sharp unanalytical change. We also study critical behavior near the transition points. Finally, we
use coherent information to quantify the scrambling of
quantum information in the system, which shows similar phase-transition-like behavior.
\end{abstract}
\maketitle
\global\long\def\ket#1{|#1\rangle}%
\global\long\def\bra#1{\langle#1|}%
\global\long\def\braket#1#2{\left\langle #1\middle|#2\right\rangle }%
\global\long\def\diff{\mathrm{d}}%
\global\long\def\inf{\infty}%
\global\long\def\pd#1#2{\frac{\partial#1}{\partial#2}}%
\global\long\def\Tr{\mathrm{Tr}}%
\renewcommand{\figurename}{FIG.}

\section{Introduction}
Chaotic quantum systems \citep{larkin1969quasiclassical,berryQuantumChaologyNot1989,maldacenaBoundChaos2016,rozenbaumLyapunovExponentOutofTimeOrdered2017,robertsChaosComplexityDesign2017}
spread initially localized information over an entire system after
isolated evolution \citep{Sekino_2008,haydenBlackHolesMirrors2007,shenkerBlackHolesButterfly2014}.
Such a process is called quantum information scrambling \citep{banulsDynamicsQuantumInformation2017,swingleUnscramblingPhysicsOutoftimeorder2018}
and lies at the heart of quantum many-body system dynamics. With the
recent development of exquisite control over multi-qubit quantum information
processing systems \citep{zhangObservationManybodyDynamical2017,aruteQuantumSupremacyUsing2019a,gongQuantumWalksProgrammable2021},
the initial information encoded in a local subsystem, which will be hidden into the whole system by quantum dynamics, can now be retrieved experimentally
by global operations \citep{miInformationScramblingComputationally2021,landsmanVerifiedQuantumInformation2019,liMeasuringOutofTimeOrderCorrelators2017,garttnerMeasuringOutoftimeorderCorrelations2017,meierExploringQuantumSignatures2019,weiExploringLocalizationNuclear2018}.
This ability can provide new insights into various fields, including
quantum chaos and quantum thermalization \citep{eisertQuantumManybodySystems2015,maldacenaBoundChaos2016,hosurChaosQuantumChannels2016},
black hole physics \citep{haydenBlackHolesMirrors2007,shenkerBlackHolesButterfly2014},
and quantum machine learning \citep{PhysRevLett.124.200504,wuArtificialNeuralNetwork2020,garciaQuantifyingScramblingQuantum2022}.

There are various methods to quantify quantum information scrambling.
One approach is to probe the spreading of an initially localized operator,
as computed by the out-of-time-ordered correlator (OTOC) \citep{robertsChaosComplexityDesign2017,hosurChaosQuantumChannels2016,swingleUnscramblingPhysicsOutoftimeorder2018,yao2016interferometric,huangOutoftimeorderedCorrelatorsManybody2017}.
It is central to the study of quantum chaos and quantum thermalization
dynamics, for its decay rate resembles the classical Lyapunov exponent
in the semi-classical limit \citep{rozenbaumLyapunovExponentOutofTimeOrdered2017}.
Also it shows the light cone structure of information propagation following the geometry
of the system \citep{swingleUnscramblingPhysicsOutoftimeorder2018,vonkeyserlingkOperatorHydrodynamicsOTOCs2018,hosurChaosQuantumChannels2016}.
Another possibility is to probe the scrambling dynamics by the correlation between
subsystems, e.g. the entanglement entropy \citep{pageAverageEntropySubsystem1993},
mutual information \citep{couchSpeedQuantumInformation2020}, and
tripartite information \citep{iyodaScramblingQuantumInformation2018}.
However, these quantities do not directly describe the amount of information
that can be extracted from a subsystem and thus may not be sufficient
to describe the dynamics of information flow.

\begin{figure}
\includegraphics[width=1\columnwidth]{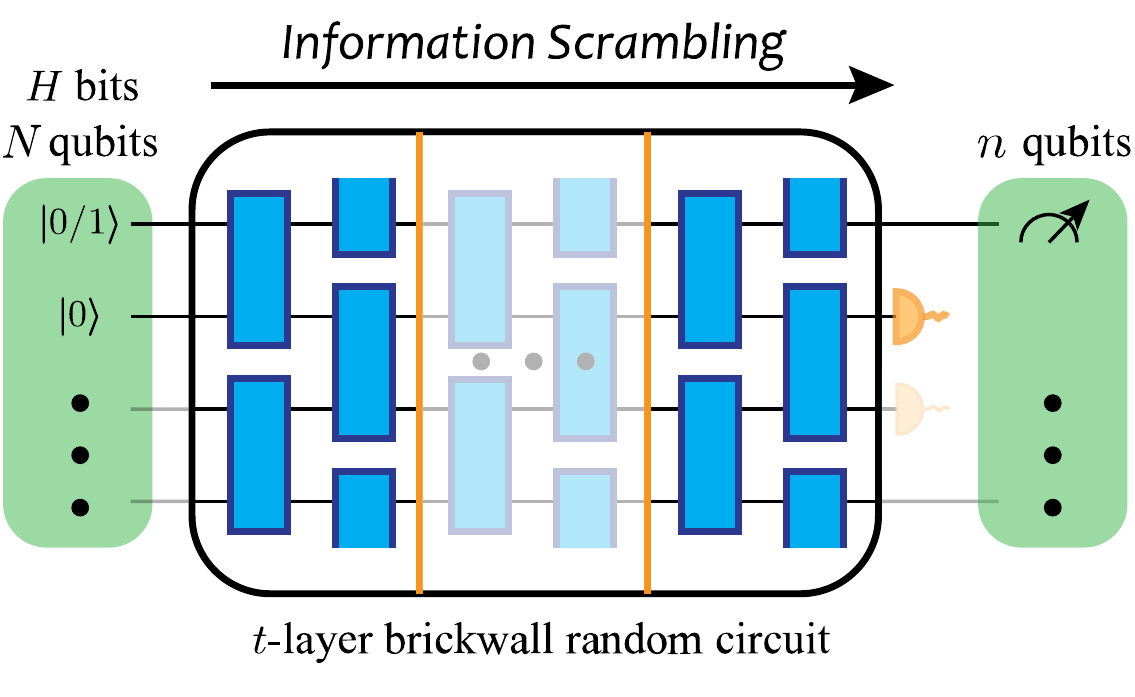}

\caption{Quantum channel scenery for probing information scrambling dynamics.
One encodes $H$-bits of information into $H$ arbitrarily chosen qubits of an $N$-qubit system. Then
after $t$ layers of \textquotedblleft brick wall\textquotedblright{}-structured random circuits, we retrieve information from a randomly selected $n$-qubit subsystem. Every \textquotedblleft brick\textquotedblright{}
(blue rectangle) represents a random unitary gate between two adjacent
qubits. \label{fig1}}
\end{figure}

To study the scrambling dynamics of quantum systems directly from
the quantum information perspective, we consider Holevo
information, which, by definition, describes the information encoded
in an ensemble of quantum states \citep{holevo1973bounds}. Since
the Holevo information is preserved under unitary evolution, it can
distinguish between the ideal case and the decoherence and thus allows us to
verify information scrambling in noisy quantum systems \citep{touilInformationScramblingDecoherence2021,landsmanVerifiedQuantumInformation2019,yoshidaDisentanglingScramblingDecoherence2019}.
Based on Holevo information, progress has been made in understanding
the distinguishability of black hole microstates in black hole theory
\citep{qiHolevoInformationEnsemble2022,baoDistinguishabilityBlackHole2017,baoHolevoInformationBlack2022}.
However, these works only focus on the final states of the black holes after
fast scrambling, while the dynamics toward scrambling is not considered.
Another related method is to apply an operator-state mapping and study the
mutual information \citep{touilInformationScramblingDecoherence2021,hosurChaosQuantumChannels2016,bertiniScramblingRandomUnitary2020}
or the tripartite information \citep{hosurChaosQuantumChannels2016}
between the input system and the output system.

Here, we consider the Hovelo information under random unitary circuits in a qubit system and show phase-transition-like behavior in the retrieved information. For long
enough circuits, as the size of the subsystem increases across a threshold of one half, the retrieved
information increases non-analytically from zero to finite values, and further increases until saturation at another transition point. This phenomenon
is observed from numerical simulation, and is also confirmed by
analytical derivation. We examine the scrambling dynamics through the convergence of the average Holevo information
toward its infinite-time limit as the circuit depth grows. We also study the critical
behavior near the phase transition points. As the Hovelo information only measures the retrieved classical information from a quantum system, finally we
use coherent information to quantify scrambling of
quantum information in the same system and find similar phase-transition-like behavior for the coherent information.

\section{Information Scrambling in Random Quantum Circuits}
Consider an $N$-qubit
quantum system. As shown in Fig.~\ref{fig1}, we store $H$ bits of classical information by randomly selecting a subsystem of $H$ qubits and preparing them into an ensemble of pure states $\left\{p_i, \ket{\psi_{i}^{\mathrm{init}}}\right\}$ with $p_{i}=\frac{1}{2^{H}}$ and ${\ket{\psi_{i}^{\mathrm{init}}}}$ be the $2^{H}$ orthogonal states in computational basis. The system then evolves
under a randomly generated Clifford circuit $U$, after which one part of
it $\mathcal{E}$ is regarded as the environment and traced out. For
the remaining system $Q$ containing $n$ qubits, we can denote the
amount of classical information that can be retrieved as $\chi_{Q}$, which is given by the Holevo information
\begin{equation}
\chi_{Q}\left(\left\{ \ket{\psi_{i}^{\mathrm{init}}}\right\} ,U\right)=S\left(\sum_{i}p_{i}\rho_{i}^{Q}\right)-\sum_{i}p_{i}S\left(\rho_{i}^{Q}\right),\label{eq:Holevo-Definition}
\end{equation}
where $\rho_{i}^{Q}=\Tr_{\mathcal{E}}(U\ket{\psi_{i}^{\mathrm{init}}}\bra{\psi_{i}^{\mathrm{init}}}U^{\dagger})$
is the output density matrix of the system $Q$, and $S$ is the von Neumann
entropy.

We adopt the periodic boundary condition for the qubits and consider random circuits
of the ``brick wall'' configuration, as shown in Fig.~\ref{fig1}. The circuit
comprises $t$ layers of alternatingly layered bricks in which each
brick represents a uniformly sampled two-qubit random Clifford gate.
We denote $\mathcal{U}_{t}$ as the set of all possible unitaries
constructed in this way with $t$ layers. Note that we choose the
Clifford circuit mainly because of the convenience in numerical simulation
\citep{aaronsonImprovedSimulationStabilizer2004a,fattalEntanglementStabilizerFormalism2004}.
Also, we specify the set of input quantum states $\left\{ \ket{\psi_{i}^{\mathrm{init}}}\right\} =\left\{ \ket{\gamma_{1}\gamma_{2}...\gamma_{H}}_{Q_{H}}\ket{0...0}_{\text{others}}\right\} _{\gamma_{i}\in\{0,1\}}$
where $Q_{H}$ is a randomly selected subsystem with $H$ qubits.
We can thus write the Holevo information as $\chi_{Q}\left(\left\{ \ket{\psi_{i}^{\mathrm{init}}}\right\} ,U\right)\equiv\chi_{Q,Q_{H}}\left(U\right)$.
Note that the choices of $Q$ and $Q_{H}$ are arbitrary and does not
need to have any specific spatial pattern.

\begin{figure}
\includegraphics[width=1\columnwidth]{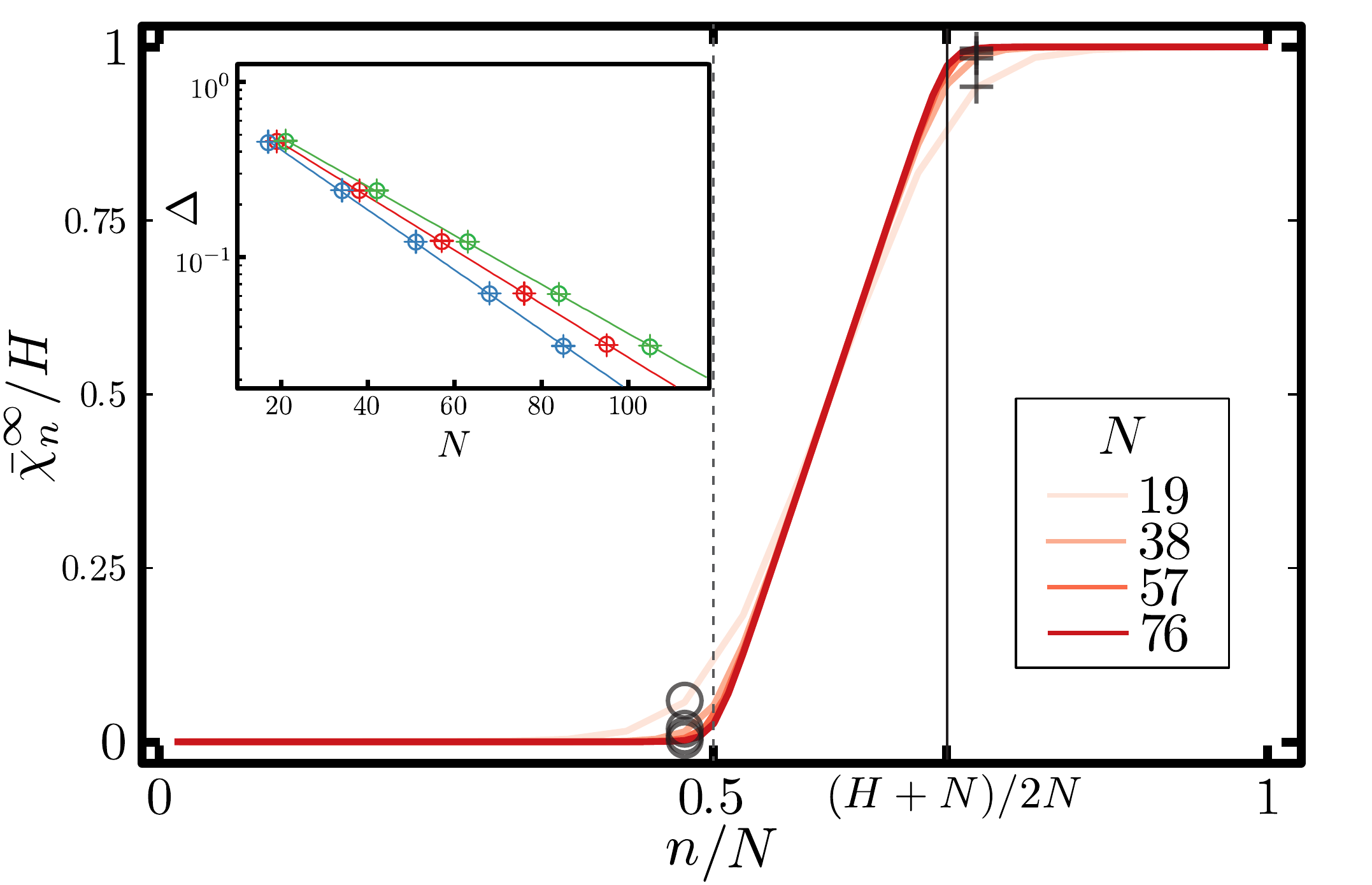}

\caption{Average Holevo information $\bar{\chi}_{n}^{\protect\inf}$ of an
$n$-qubit subsystem $Q$ in the steady state. For the later convenience
to examine the behavior near the phase transition point, we
choose $N$ to be multiples of $19$. For different $N$ we fix its
ratio with the encoded classical information as $N:H=19:8$. The curve
becomes sharper around the two phase transition points
as $N$ increases. Note that the behavior is similar for generic
$1\le H\protect\leq N$. The inset shows the result of finite-size
scaling. We focus on the difference $\Delta$ between $\bar{\chi}_{n}^{\protect\inf}$
and the thermodynamic limit $\chi_{n}^{\text{thermal}}$ when $n$ is chosen to be close to
the phase transition point $\frac{n}{N}=\frac{1}{2}$ (circle) and $\frac{n}{N}=\frac{H+N}{2N}$
(cross). Specifically, for $n$ smaller
than $\frac{N}{2}$ we calculate $\Delta_{1}=\bar{\chi}_{n}^{\protect\inf}$, and for $n$ larger than $\frac{N+H}{2}$ we calculate $\Delta_{2}=H-\bar{\chi}_{n}^{\protect\inf}$. The blue, red, and green lines are for three different ratios
of $N:H=17:6,\allowbreak19:8,\allowbreak21:10$, respectively.
When $n,N,H$ increase under a fixed ratio, $\Delta_{1}$ and $\Delta_{2}$
decay exponentially. For each data point, we sample
$10^{6}$ times. The long-time limit is approximated
by choosing $t=3N$ as described in the main text.\label{fig2}}
\end{figure}

After obtaining the Holevo information contained in a randomly selected
subsystem from the above setting, we further average over all possible
subsystem $Q$, all possible input states, and all the circuits with
the same depth $t$ to get
\[
\bar{\chi}_{n}^{t}=\frac{1}{|\mathcal{U}_{t}|}\frac{1}{|\mathcal{S}_{n}|}\frac{1}{|\mathcal{S}_{H}|}\sum_{U\in\mathcal{U}_{t}}\sum_{Q\in\mathcal{S}_{n}}\sum_{Q_{H}\in\mathcal{S}_{H}}\chi_{Q,Q_{H}}\left(U\right),
\]
where $\mathcal{S}_{k}$ denotes the set of all subsystems with $k$ qubits.

\begin{figure*}
\includegraphics[width=2\columnwidth]{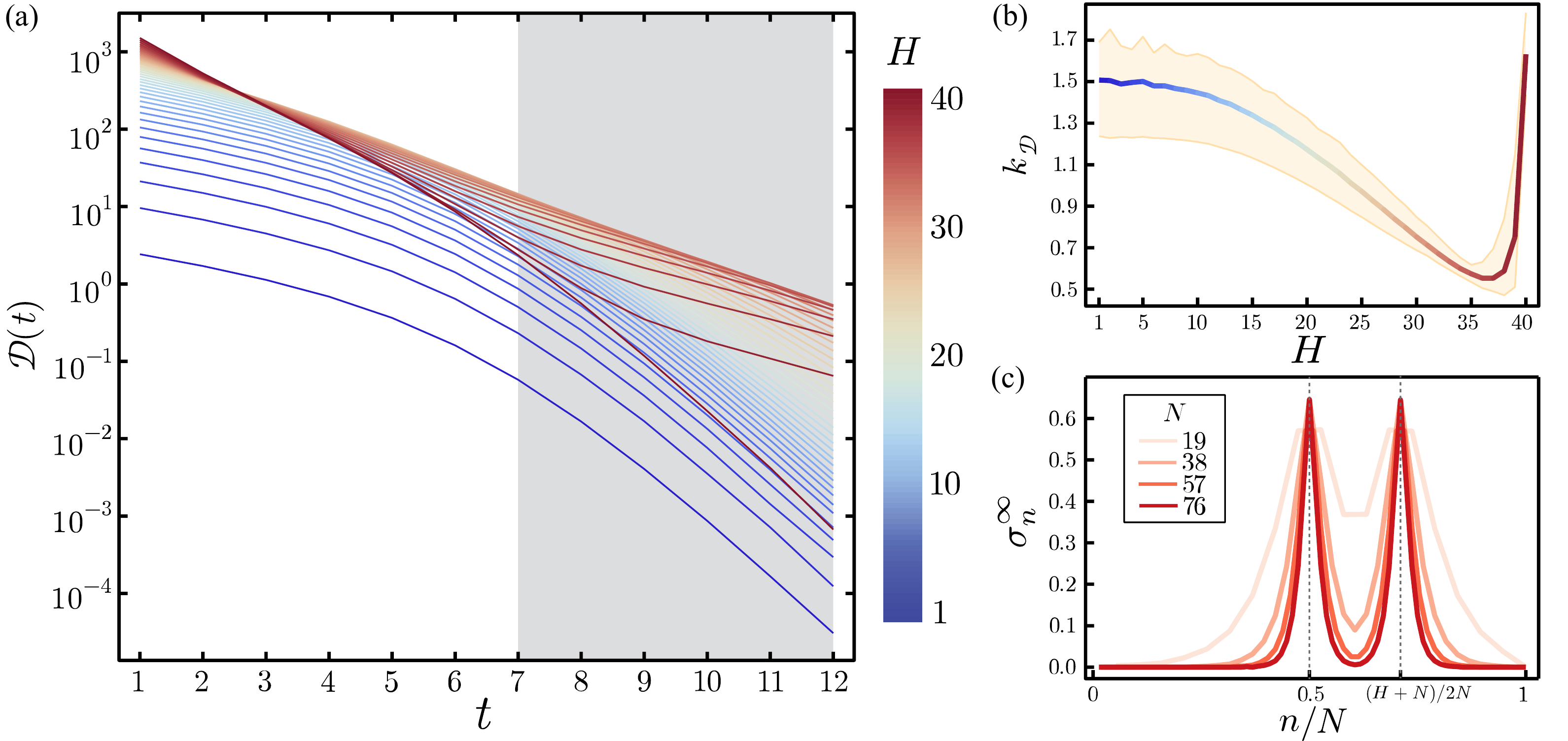}

\caption{(a) Distance $\mathcal{D}(t)$ of the averaged Holevo information
from its long time limit. We calculate for $N=40$ and $H$ from $1$
(blue) to $N$ (red). Each data point is obtained by averaging over
$10^{6}$ random circuits and initial states. (b) The decay rate of
$\mathcal{D}(t)$ in the greyed region of (a), defined as $k_{\mathcal{D}}^{7,12}$
in the main text. The shaded region around the curve shows the confidence interval
whose upper and lower bounds are estimated by the maximum and minimum
slopes in the greyed region in (a). (c) Standard deviation $\sigma_{n}^{\protect\inf}$
of Holevo information $\chi_{Q}(U)$ in $n$ qubit subsystem $Q\in\mathcal{S}_{n}$.
The steady-state is approximated by setting circuit depth $t=3N$.
We fix the ratio $N:H=19:8$. At the two phase transition points $\frac{n}{N}=\frac{1}{2},\frac{H+N}{2N}$,
the peaks become sharper as $N$ grows. \label{fig:fig3}}
\end{figure*}

As shown in Fig. \ref{fig2}, we numerically compute the long time limit of the
Holevo information $\bar{\chi}_{n}^{\inf}=\lim_{t\rightarrow\inf}\bar{\chi}_{n}^{t}$
by setting $t=3N$. Theoretically, $\Omega(N)$ layers are needed for the initially localized information to propagate over the whole system \citep{haydenBlackHolesMirrors2007,nahumQuantumEntanglementGrowth2017,hunter-jonesUnitaryDesignsStatistical2019a}, and here, we verify the convergence of Holevo information by comparing the results at $t=3N$ with those at $t=4N$.
A phase-transition-like behavior can be observed: For small system size $n$, we
are not able to retrieve any information; When $n$ reaches half of
the system size, information starts emerging at a constant rate of
two bits of classical information per qubit; Finally, the retrievable information
reaches its maximum value through another sharp nonanalytical change, indicating
that all of the initially encoded information can be reliably recovered. We
perform finite-size scaling near the two points to further analyze the phase-transition-like behavior. As shown in the inset of Fig.~\ref{fig2}, by fixing
a ratio at $\frac{n}{N}<\frac{1}{2}$ and $\frac{n}{N}>\frac{H+N}{2N}$,
respectively, and increasing $n,N,H$ simultaneously, $\bar{\chi}_{n}^{\inf}$
converges exponentially towards its thermodynamic limit. Note that
one can get the same value of $\bar{\chi}_{n}^{\inf}$ without averaging
over the choice of $Q_{H}$. This can be understood from the definition
of scrambling and from the symmetry of the random unitary group \citep{pageAverageEntropySubsystem1993}.

Indeed, if the circuit is sampled uniformly over the $N$-qubit Clifford
group, we can calculate theoretically the average Holevo information
$\bar{\chi}_{n}^{\inf}$ for arbitrary $n,N,H$, and it agrees well
with the numerical result obtained above for layered random two-qubit Clifford gates. For more details, see Appendix \ref{Asec}. Specifically, if we take the thermodynamic limit $n, N, H\rightarrow \infty$ with the ratio $\frac{n}{N} \equiv r_n , \frac{H}{N} \equiv r_H $ held constant, this theoretical value $\bar{\chi}_{n}^{\inf}$
converges to 
\begin{align}
\frac{1}{H}\chi_{n}^{\text{thermal}} & =\begin{cases}
0 & r_n\leq\frac{1}{2}\\
(2r_n-1)/r_H & \frac{1}{2}<r_n\leq\frac{1}{2}(1+r_H)\\
1 & r_n>\frac{1}{2}(1+r_H)
\end{cases}.\label{eq:thermal}
\end{align}

This further allows us to define the critical exponent
\[
k\equiv\lim_{\tau\rightarrow0^{+}}\frac{\log\left|f(\tau)\right|}{\log\left|\tau\right|}=1
\]
where for the first transition point $\tau=r_n-\frac{1}{2}$,
$f(\tau)=\frac{1}{H}\chi_{n}^{\text{thermal}}$ and similarly for the second transition point.

\section{Information Scrambling Dynamics} 
From the evolution of the Holevo information, we can study the information scrambling dynamics
of quantum systems. For example, here we study how the long-time limit of the average Holevo information is approached.
We compare the average Holevo information for different subsystem
sizes $\left\{ \bar{\chi}_{n}^{t}\right\} _{n=1}^{N}$ with its long-time limit. We use 2-norm to measure their difference
\[
\mathcal{D}(t)=\sum_{n=1}^{N}(\bar{\chi}_{n}^{t}-\bar{\chi}_{n}^{\inf})^{2}.
\]

From the numerical simulation, we plot $\mathcal{D}(t)$ for the system
size $N=40$ and $H$ ranging from $1$ to $N$, as shown in Fig.~\ref{fig:fig3}a. We observe that $\left\{ \bar{\chi}_{n}^{t}\right\} _{n=1}^{N}$
converges under time evolution roughly exponentially. Further, the
decaying rate, which corresponds to the scrambling speed, varies for different initial Holevo information $H$.

To compare the speed of scrambling between $1\leq H\leq N$, we extract the average slope on the semi-log plot of $\mathcal{D}(t)$ between time $t$ and $t^\prime$ as
\[
k_{\mathcal{D}}^{t,t^{\prime}}=\left|\frac{\log\mathcal{D}(t)-\log\mathcal{D}(t^{\prime})}{t-t^{\prime}}\right|
\]
as shown in Fig.~\ref{fig:fig3}b. The upper and lower confidence bounds are roughly
estimated by the maximum and minimum slope in the region. For the same set of circuits, the
scrambling rate slows down when an increasing amount of information
is encoded into the system until close to $\frac{H}{N}\sim1$ where
the tendency reverses, which may be caused by the finite size effect.

We further study the information scrambling behavior of individual realizations $U$ of random
circuits by comparing the Holevo information distribution $\chi_{Q,Q_{H}}\left(U\right)$
with the average value over different realizations. We characterize it by the standard
deviation $\sigma_{n}^{t}$
\[
\left(\sigma_{n}^{t}\right)^{2}=\frac{1}{|\mathcal{U}_{t}|}\frac{1}{|\mathcal{S}_{n}|}\frac{1}{|\mathcal{S}_{H}|}\sum_{U\in\mathcal{U}_{t}}\sum_{Q\in\mathcal{S}_{n}}
\sum_{Q_{H}\in\mathcal{S}_{H}}(\chi_{Q,Q_{H}}\left(U\right)-\bar{\chi}_{n})^{2}.
\]

As shown in Fig.~\ref{fig:fig3}c, we numerically compute its long time limit $\sigma_{n}^{\inf}=\lim_{t\rightarrow\inf}\sigma_{n}^{t}$
by setting $t=3N$. The ratio $N:H=19:8$ is set in accordance with
Fig.~\ref{fig2}. As we can see, $\sigma_{n}^{\inf}$ is asymptotically zero not
only in the region $\frac{n}{N}<\frac{1}{2}$ and $\frac{n}{N}>\frac{N+H}{2N}$,
where $\bar{\chi}_{n}^{\inf}$ already saturates, but also in the
region $\frac{1}{2}<\frac{n}{N}<\frac{N+H}{2N}$. This suggests that
information almost fully scrambles even for a single typical
random circuit. Only at the two phase transition points can we get finite standard deviations, which do not increase with $N$, $H$ and $n$ so the relative fluctuation is decreasing
for larger systems.

\section{Similar Phase-Transition-Like Behavior for Coherent Information}
After studying the information scrambling dynamics by the classical information, a natural next step is to consider if the same scrambling dynamics can be probed by quantum information as well, particularly if the same phase-transition-like behavior persists. 

Coherent information \citep{schumacherQuantumDataProcessing1996,lloydCapacityNoisyQuantum1997,devetakClassicalDataCompression2003}
quantifies the remaining quantum information after a state goes through
a quantum channel, with similar properties as the mutual information
in classical communication. The coherent information is also related
to the reversibility of the quantum channel \citep{holevoQuantumChannelsTheir2012}
and the condition of quantum error correction \citep{schumacherQuantumDataProcessing1996},
thus lies at the heart of understanding the difference between classical
and quantum information communication.

\begin{figure}
\includegraphics[width=1\columnwidth]{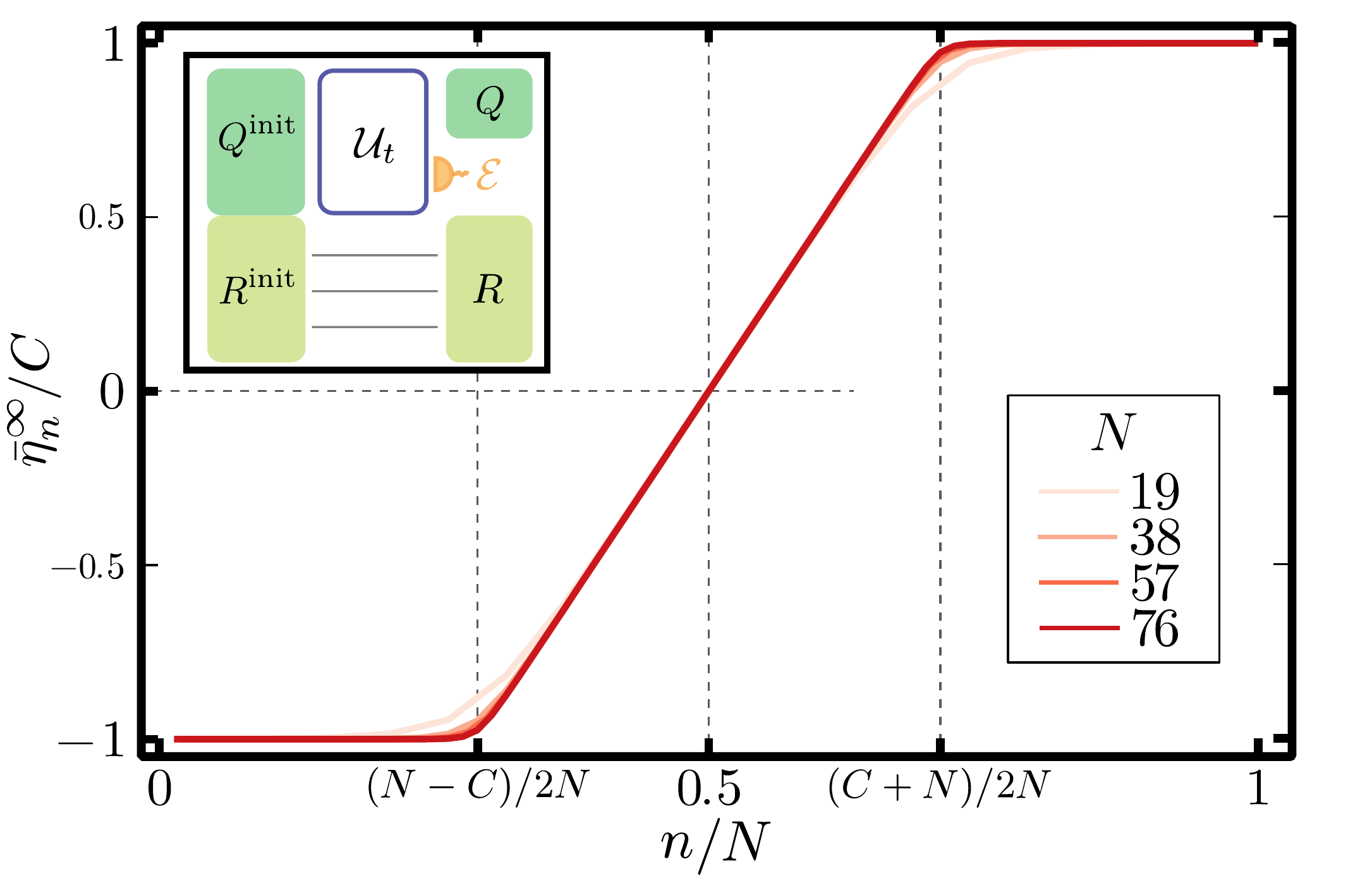}

\caption{Average coherent information $\bar{\eta}_{n}^{\protect\inf}$ of an
$n$-qubit subsystem $Q$ in the steady state. Like the settings for
the Holevo information, we scale the system size $N$ and the total
encoded quantum information $C$ by $N:C=19:8$. As $N$ grows,
the curve shows two phase transition points
at $\frac{n}{N}=\frac{N-C}{2N},\allowbreak\frac{n}{N}=\frac{N+C}{2N}$,
respectively. The inset shows an illustration of the numerical scheme.
The initial state, encoded as an ensemble into $Q^{\text{init}}$,
is purified by $R^{\text{init}}$. After a random circuit $U\in\mathcal{U}_{t}$,
part of the system $\mathcal{E}$ is traced out as the environment.
Together they form a quantum channel $\mathcal{C}$. The coherent
information encoded in the system $Q$ can thus be computed with the
help of the reference system $R$. \label{fig4}}
\end{figure}

As shown in the inset of Fig.~\ref{fig4}, similar to the model for Holevo information,
we encode quantum information of $C$ in $N$
qubits with periodic boundary conditions, apply random Clifford circuit $U\in\mathcal{U}_{t}$ and regard a
randomly selected subsystem $\mathcal{E}$ as the environment
to be traced off. The difference is that the system's initial state
$Q^{\text{init}}$ would be an ensemble $\rho^{\mathrm{init}}=\frac{1}{2^{C}}I_{Q_{C}}\otimes\ket{0...0}\bra{0...0}_{\text{others}}$,
where $Q_{C}$ represents a subsystem of randomly selected $C$ qubits.
This ensemble can be seen as a mixture of $\left\{ \ket{\psi_{i}^{\mathrm{init}}}\right\} $
with equal probabilities. We write the final state of the $n$-qubit
system $Q$ as $\rho^{Q}$. The circuit, together with tracing out the environment
$\mathcal{E}$, forms a quantum channel $\mathcal{C}$. The coherent information
can thus be calculated as \citep{schumacherQuantumDataProcessing1996,lloydCapacityNoisyQuantum1997,devetakClassicalDataCompression2003}
\[
\eta_{Q}(\rho^{\mathrm{init}},\mathcal{C})=S(\rho^{Q})-S(\rho,\mathcal{C}),
\]
where $S(\rho,\mathcal{C})$ is the entropy exchange. By definition,
to calculate $S(\rho,\mathcal{C})$, we need to purify $Q^{\text{init}}$
using a reference system $R^{\text{init}}$ before applying the quantum
channel. Then we get $S(\rho,\mathcal{C})=S(\rho^{QR})$.

Similar to what we have done for Holevo information, we write $\eta_{Q}(\rho^{\mathrm{init}},\mathcal{C})\equiv\eta_{Q,Q_{H}}(\mathcal{C})$
and average over the system $Q$, the input states and the circuit
to get
\[
\bar{\eta}_{n}^{t}=\frac{1}{|\mathcal{U}_{t}|}\frac{1}{|\mathcal{S}_{n}|}\frac{1}{|\mathcal{S}_{C}|}\sum_{U\in\mathcal{U}_{t}}\sum_{Q\in\mathcal{S}_{n}}\sum_{Q_{H}\in\mathcal{S}_{C}}\eta_{Q,Q_{H}}(\mathcal{C}).
\]
Finally, the long-time limit $\bar{\eta}_{n}^{\inf}$ is approximated
by $t=3N$.

The results for various system sizes are shown in Fig.~\ref{fig4} which has
a phase-transition-like behavior similar to that of the Holevo information,
although the phase transition points are at $\frac{n}{N}=\frac{N-C}{2N},\frac{n}{N}=\frac{N+C}{2N}$.
To see how these transition points correspond to those for the Holevo information, note that the second transition point $\frac{n}{N}=\frac{N+C}{2N}$ is the same for both cases. On the other hand, when $\frac{n}{N}<\frac{N-C}{2N}$, all the information goes into the environment,
making the coherent information saturate to its lower bound $\frac{\bar{\eta}_{n}^{\inf}}{C}=-1$.
The phase transition point at $\frac{n}{N}=\frac{N-C}{2N}$ thus corresponds
to that of the Holevo information in the environment $\mathcal{E}$ rather than in the system $Q$.
Finally, when $n<\frac{N}{2}$ we have negative $\bar{\eta}_{n}^{\inf}$, indicating
that no quantum information can be retrieved from the output system. This is in agreement with
the result for Holevo information.

\section{Discussions}
In summary, in this work we use the spatial distribution of
Holevo information to characterize the information scrambling process.
The information converges to zero in the thermodynamic limit when
we consider subsystem sizes smaller than half the system. When exceeding
this threshold, the extractable classical information increases by two bit per added
qubit until its saturation to the total encoded information.
This can serve as a scrambling criterion, and its comparison with
others, including Haar scrambled \citep{haydenBlackHolesMirrors2007}
and Page scrambled \citep{Sekino_2008} criteria, is of great interest. We
study how the system approaches the long-time limit and how the convergence speed varies with the amount of encoded information.
We also find that variation around the average behavior is vanishingly small almost everywhere apart from the two phase transition points, which implies that in the thermodynamic limit, almost all random circuits
would meet the scrambling criteria. Finally, we find that the coherent
information possesses a similar phase-transition-like behavior.

One can regard the discarded environment in our model as the
qubit loss error from the quantum error correction (QEC) perspective
\citep{aharonovFaulttolerantQuantumComputation1997,kitaevQuantumComputationsAlgorithms1997}.
Thus the phase transition point $\frac{n}{N}=\frac{N+C}{2N}$ of coherent
information would correspond to the condition of perfect decoding.
Specifically, our model uses the random circuit to encode $C$ logical
qubits in $N$ physical qubits. This code can tolerate the loss of
$\frac{N-C}{2}$ located qubits which saturates the quantum Singleton
bound \citep{knillTheoryQuantumErrorcorrecting1997}.

Although here we restrict the calculation to Clifford gates for numerical
convenience, this method using Holevo information to characterize information scrambling should largely be
applicable to generic quantum systems. Specifically, this process of encoding information
by a set of initial states and calculating the Holevo information of
a selected subsystem in the final states does not require any special property
of the intermediate quantum dynamics. We can thus easily extend
the unitary evolution to arbitrary quantum channels, although the detailed late time physics may depend on specific models and remains an open direction for future research. Therefore it
may provide a universal tool for probing quantum information scrambling
dynamics.

\section*{Acknowledgments}
We thank Z.-D. Liu, D. Yuan and T.-R. Gu for discussions. This work was supported by the Frontier Science Center for Quantum Information of the Ministry of Education of China and the Tsinghua University Initiative Scientific Research Program.

\appendix

\renewcommand{\appendixname}{APPENDIX}
\newcommand{\appsection}[1]{\section{\MakeUppercase{#1}}}
\renewcommand{\thesubsection}{\arabic{subsection}}
\makeatletter
\renewcommand{\p@subsection}{\thesection. }
\makeatother
\appsection{Proof of Eq. (\ref{eq:thermal})\label{Asec}}
\subsection{Model Description}

Consider an $n$-qubit subsystem $Q$ in an $N$-qubit Clifford system.
In the main text, we have already specified that the intitial quantum
states are $\left\{ \ket{\psi_{i}^{\mathrm{init}}}\right\} =\left\{ \ket{\gamma_{1}\gamma_{2}...\gamma_{H}}_{Q_{H}}\ket{0...0}_{\text{others}}\right\} _{\gamma_{i}\in\{0,1\}}$
each appearing with equal probability $\frac{1}{2^{H}}$. After a
random Clifford circuit $U$, we denote the final state as $\ket{\psi_{i}}$.
The Holevo Information can be written as $\chi_{n}=S_{n}(U\frac{1}{2^{H}}\sum_{2^{H}}\ket{\psi_{i}}\bra{\psi_{i}}U^{\dagger})-\frac{1}{2^{H}}\sum_{2^{H}}S_{n}(U\ket{\psi_{i}}\bra{\psi_{i}}U^{\dagger})$,
where $S_{n}(\rho)$ represents the entropy of the $n$-qubit subsystem
$Q$. 

When we average it over all possible circuits with its depth large
enough, the first and the second term converges respectively. Thus,
we can decompose it into two parts: 
\begin{equation}
\bar{\chi}_{n}=\mathbb{E}S_{n,H}-\mathbb{E}S_{n,0}\label{eq:Holevo}
\end{equation}
 where we write $\rho_{h}=\frac{1}{2^{h}}\sum_{2^{h}}\ket{\psi_{i}}\bra{\psi_{i}}$
and $S_{n,h}=S_{n}(U\rho_{h}U^{\dagger})$ for convenience.

The Clifford unitary with large depth will bring any initial state
$\rho_{h}$ into a finite set $\mathrm{Orb}(\rho_{h},\mathcal{U}^{N})$
with uniform probability distribution, where $\mathrm{Orb}(\rho_{h},\mathcal{U}^{N})$
is the orbit of $\rho_{h}$ under the $N$-qubit Clifford group $\mathcal{U}^{N}$.
Thus the probability $\text{Prob}(S_{n,h}=x)$ would be proportional
to the number of states $\rho\in\mathrm{Orb}(\rho_{h})$ that satisfies
$S_{n}(\rho)=x$. Here we calculate the number of elements in such
set $\left|\left\{ \rho\in\mathrm{Orb}(\rho_{h},\mathcal{U}^{N})|S_{n}(\rho)=x\right\} \right|$
and give the expectation value $\mathbb{E}S_{n,h}$.

\pagebreak
\begin{widetext}
\subsection{Outline of the Proof}

Any state $\rho_{h}$ that satisfies $S(\rho_{h})=x$ can be transformed
by a local Clifford unitary $U\in G=\left\{ U(n)\otimes U(m)\right\} $
to $\rho_{h}^{x}$ which we write under the stabilizer formalism \citep{fattalEntanglementStabilizerFormalism2004,aaronsonImprovedSimulationStabilizer2004a}

\begin{equation}   
\begin{bmatrix}[cccc|cccc]
X &   &   &   & X &     \\
Z &   &   &   & Z &      \\
  & Z &   &   &   & Z    \\
  &   & Z &   &   &   & Z\\
&  &  & Z & \\ 
&  &  &  &  &  &  & Z
\end{bmatrix}    
\begin{BMAT}(@)[1pt,10pt,0pt]{l}{cccc}     
\scalebrace{2}{k_1 \text{ pairs}}\\     
\scalebrace{2}{k_2 }\\      
\scalebrace{1}{l_1}\\     
\scalebrace{1}{l_2}\\    
\end{BMAT} 
\label{ass}
\end{equation}where $n+m=N$, $x=n-l_{1}$ and the number of stabilizer operators
\begin{equation}
2k_{1}+k_{2}+l_{1}+l_{2}=n+m-h\label{eq:num operators}
\end{equation}

We can constraint the four parameters by 
\begin{equation}
\left\{ \begin{aligned}k_{1}+k_{2}+l_{1} & \leq n\\
k_{1}+k_{2}+l_{2} & \leq m\\
k_{1},k_{2},l_{1},l_{2} & \geq0
\end{aligned}
\right.\label{eq:constraints}
\end{equation}

We can further simplify our question by 
\[
\left\{ \rho\in\mathrm{Orb}(\rho_{h},\mathcal{U}^{N})|S_{n}(\rho)=x\right\} =\left\{ \rho\in\mathrm{Orb}(\rho_{h},G)|S_{n}(\rho)=x\right\} =\left\{ \rho\in\mathrm{Orb}(\rho_{h}^{x},G)\right\} 
\]
This can be given by  directly using the common state Eq. \ref{ass}
that all the states with the same entropy can reach by local unitaries. 

With the help of Lagrange's orbit-stabilizer theorem $\left|\mathrm{Orb}(\rho_{h}^{x},G)\right|=\frac{\left|G\right|}{\left|\mathrm{Stab}(\rho_{h}^{x},G)\right|}$,
we only need 
\begin{itemize}
\item the order of unitary group $\left|G\right|$ 
\item the order of stabilizer $\left|\mathrm{Stab}(\rho_{h}^{x},G)\right|$.
Here the stabilizer $\mathrm{Stab}(\rho_{h}^{x},G)$ represents the
set of elements in $G$ that makes $\rho_{h}^{x}$ invariant.
\end{itemize}
We can get $\left|G\right|$ directly from the volume of $N$ qubit
Clifford group $Cl(N)=\prod_{j=1}^{N}2(4^{j}-1)4^{j}$ \citep{nebeInvariantsCliffordGroups2000}
\[
\left|G\right|=Cl(n)\times Cl(m)
\]

Based on basic combinatorics, we have from Sec. \ref{sec:Proof-of-Combinatorics}

\begin{align}
\left|\mathrm{Stab}(\rho_{h}^{x},G)\right|= & \left(\prod_{j=1}^{k_{1}}\left(4^{j}-1\right)4^{j}2\right)2^{2k_{1}(k_{2}+l_{1})}2^{2k_{1}l_{2}}\label{eq:combinatorics}\\
\times & 2^{k_{2}}2^{2k_{2}}2^{\frac{1}{2}k_{2}(k_{2}+2l_{1}+1)}2^{\frac{1}{2}k_{2}(k_{2}+2l_{2}+1)}4^{h_{1}k_{2}}4^{h_{2}k_{2}}\left(\prod_{j=0}^{k_{2}-1}\left(2^{k_{2}}-2^{j}\right)\right)2^{l_{1}k_{2}}2^{l_{2}k_{2}}\nonumber \\
\times & 2^{l_{1}+\frac{1}{2}l_{1}(l_{1}+1)}\left(\prod_{j=0}^{l_{1}-1}\left(2^{l_{1}}-2^{j}\right)\right)4^{h_{1}l_{1}}\nonumber \\
\times & 2^{l_{2}+\frac{1}{2}l_{2}(l_{2}+1)}\left(\prod_{j=0}^{l_{2}-1}\left(2^{l_{2}}-2^{j}\right)\right)4^{h_{2}l_{2}}\nonumber \\
\times & Cl(h_{1})Cl(h_{2})\nonumber 
\end{align}
where 
\begin{align*}
h_{1} & =n-k_{1}-k_{2}-l_{1}\\
h_{2} & =m-k_{1}-k_{2}-l_{2}
\end{align*}

When $n,h,N$ is fixed, the order of the orbit can be determined by
the four parameters $k_{1},k_{2},l_{1},l_{2}$.
\[
\left|\mathrm{Orb}(\rho_{h}^{x},G)\right|=\frac{\left|G\right|}{\left|\mathrm{Stab}(\rho_{h}^{x},G)\right|}\equiv t_{k_{1},k_{2},l_{1},l_{2}}
\]

Recall that $x=n-l_{1}$, we have
\begin{equation}
\mathbb{E}S_{n,h}=\frac{\sum(n-l_{1})\times t_{k_{1},k_{2},l_{1},l_{2}}}{\sum t_{k_{1},k_{2},l_{1},l_{2}}}\label{eq:Expectation_Snh}
\end{equation}
where the sum is over all possible $k_{1},k_{2},l_{1},l_{2}$ under
the constraint Eq. (\ref{eq:constraints}).

By setting $h=0$ and $h=H$, this gives both the terms in Eq. (\ref{eq:Holevo}).
For specific values of $n,H,N$, the calculation agrees well with
our numerical result in the main text, as shown in Fig. \ref{fig:Numerical-simulation-using}

\begin{figure}
\includegraphics[width=1\columnwidth]{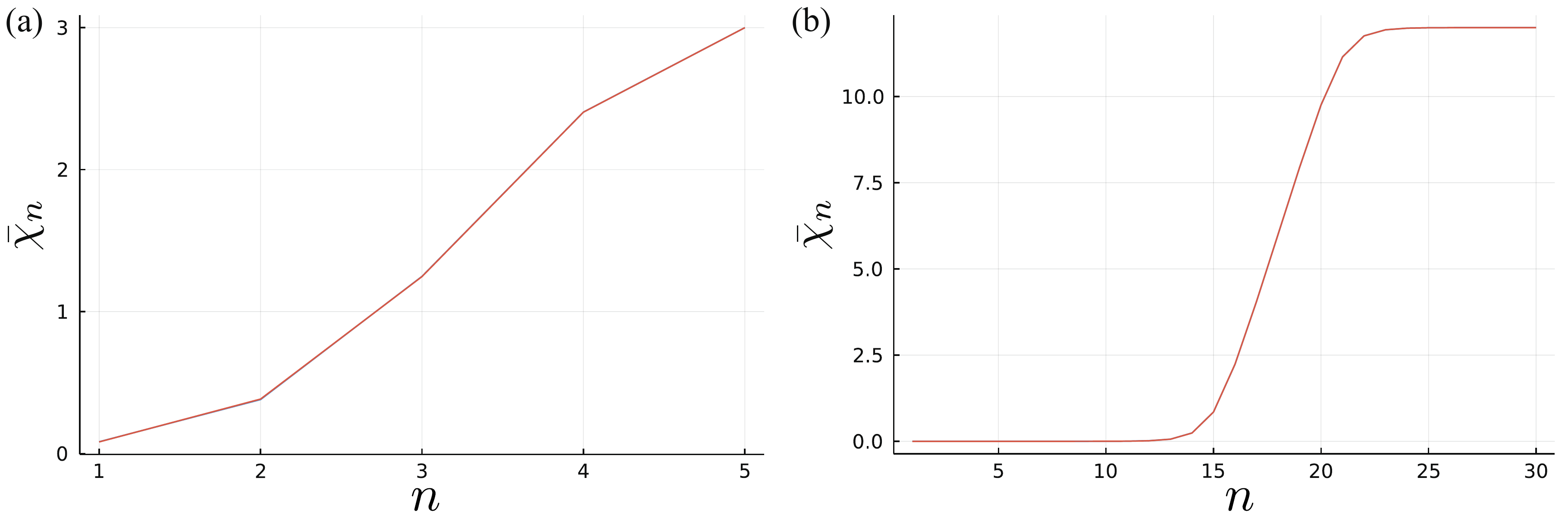}

\caption{Numerical simulation using Monte Carlo method (red) and analytic calculation
(blue) of Holevo information $\bar{\chi}_{n}$. (a) $N=5$ and $H=3$.
(b) $N=30$ and $H=15$. The two lines almost completely overlap.\label{fig:Numerical-simulation-using}}
\end{figure}

To further prove Eq. (\ref{eq:thermal}) in the main text, we first observe that $t_{k_{1},k_{2},l_{1},l_{2}}$ is varying in
an exponential way for different $k_{1},k_{2},l_{1},l_{2}$. When
$n,N,H$ is large, the expected value of $n-l_{1}$ would correspond
to $k_{1},k_{2},l_{1},l_{2}$ where $t_{k_{1},k_{2},l_{1},l_{2}}$
reaches maximum.

Assuming $k_{1},k_{2},l_{1},l_{2}\rightarrow\inf$ and ignoring constant
terms, we can convert the question of where $\log_{2}t_{k_{1},k_{2},l_{1},l_{2}}$
reaches maximum to be where $f(k_{1},k_{2},l_{1},l_{2})$ reaches
maximum, here 

\begin{align*}
-f(k_{1},k_{2},l_{1},l_{2}) & =2k_{1}(N-H-k_{1})+3k_{1}\\
 & +3k_{2}+2k_{2}(k_{2}+l_{1}+l_{2}+H)\\
 & +l_{1}\left(\frac{3}{2}l_{1}+\frac{1}{2}\right)+l_{2}\left(\frac{3}{2}l_{2}+\frac{1}{2}\right)\\
 & +2(n-k_{1}-k_{2}-l_{1})(n-k_{1}-k_{2}+1)\\
 & +2(N-n-k_{1}-k_{2}-l_{2})(N-n-k_{1}-k_{2}+1)
\end{align*}

Combining with the constraints in Eq. (\ref{eq:num operators}) and
Eq. (\ref{eq:constraints}), $f(k_{1},k_{2},l_{1},l_{2})$ reaches
maximum when 

\begin{equation}
\begin{array}{c}
n<\frac{N-h}{2}\\
\\
\left\{ \begin{aligned}k_{1} & =n\\
k_{2} & =0\\
l_{1} & =0\\
l_{2} & =N-h-2n
\end{aligned}
\right.
\end{array}\begin{array}{c}
\frac{N-h}{2}\leq n\leq\frac{N+h}{2}\\
\\
\left\{ \begin{aligned}k_{1} & =(N-h)/2\\
k_{2} & =0\\
l_{1} & =0\\
l_{2} & =0
\end{aligned}
\right.
\end{array}\begin{array}{c}
n>\frac{N+h}{2}\\
\\
\left\{ \begin{aligned}k_{1} & =N-n\\
k_{2} & =0\\
l_{1} & =2n-N-h\\
l_{2} & =0
\end{aligned}
\right.
\end{array}\label{eq:maximum}
\end{equation}
the proof is given in Sec. \ref{sec:Proof-of-Eq.Maximum}

By Eq. (\ref{eq:Expectation_Snh}) we get 
\[
\mathbb{E}S_{n,h}=\min(n,N+h-n)
\]

Finally the Holevo information 
\[
\bar{\chi}_{n}=\mathbb{E}S_{n,H}-\mathbb{E}S_{n,0}=\begin{cases}
0 & n<\frac{N}{2}\\
2n & \frac{N-H}{2}\leq n\leq\frac{N+H}{2}\\
H & n>\frac{N+H}{2}
\end{cases}
\]

\subsection{Proof of Eq. (\ref{eq:combinatorics}) \label{sec:Proof-of-Combinatorics}}

We count the number of elements in $G$ that makes $\rho_{h}^{x}$
invariant. Every non-trivial elements in the $G=\left\{ U(n)\otimes U(m)\right\} $
maps the set of stabilizer operators in Eq. \ref{ass} to a new one.
We count how many of the resulting stabilizer operators form a equivalent
state to the original. Two sets of stabilizers are equivalent if they
are the same under standard stabilizer multiplication operations.

Below we split Eq. (\ref{eq:combinatorics}) into factors and explain
them respectively.
\begin{itemize}
\item $\left(\prod_{j=1}^{k_{1}}\left(4^{j}-1\right)4^{j}2\right)2^{2k_{1}(k_{2}+l_{1})}2^{2k_{1}l_{2}}$.
For the first $2k_{1}$ stabilizers, the two in each pair do not commute
with each other in two subsystems. This property would preserve for
arbitrary local transformation. 
\item $2^{k_{2}}2^{2k_{2}}2^{\frac{1}{2}k_{2}(k_{2}+2l_{1}+1)}2^{\frac{1}{2}k_{2}(k_{2}+2l_{2}+1)}4^{h_{1}k_{2}}4^{h_{2}k_{2}}\left(\prod_{j=0}^{k_{2}-1}\left(2^{k_{2}}-2^{j}\right)\right)2^{l_{1}k_{2}}2^{l_{2}k_{2}}$.
For the next $k_{2}$ stabilizers, the combination of arbitrary elements
in $l_{1}+l_{2}$ stabilizers are reachable by $G$.
\item $2^{l_{1}+\frac{1}{2}l_{1}(l_{1}+1)}\left(\prod_{j=0}^{l_{1}-1}\left(2^{l_{1}}-2^{j}\right)\right)4^{h_{1}l_{1}}$.
The local unitaries in $G$ can only bring $l_{1}$ stabilizers which
is local in $Q$ to another local stabilizer in $Q$.
\item $2^{l_{2}+\frac{1}{2}l_{2}(l_{2}+1)}\left(\prod_{j=0}^{l_{2}-1}\left(2^{l_{2}}-2^{j}\right)\right)4^{h_{2}l_{2}}$.
The same as $l_{1}$
\item $Cl(h_{1})$. After determining all the stabilizers above, there still
exists $h_{1}$ degrees of freedom in $Q$. This corresponds to $Cl(h_{1})$
elements.
\item $Cl(h_{2})$. The same as $h_{1}$.
\end{itemize}

\subsection{Proof of Eq. (\ref{eq:maximum}) \label{sec:Proof-of-Eq.Maximum}}

$f(k_{1},k_{2},l_{1},N-h-(2k_{1}+k_{2}+l_{1}))$ is a concave function
for $k_{1},k_{2},l_{1}$. This can be seen from its Hessian matrix
$\left(\begin{array}{ccc}
-8 & -4 & -6\\
-4 & -7 & -3\\
-6 & -3 & -6
\end{array}\right)$

At Eq. (\ref{eq:maximum}), we only need to prove that $f$ achieves
its local maximum to prove that it also achieves global maximum over
the convex region defined in Eq. (\ref{eq:constraints}). Further,
we observe that the condition in Eq. (\ref{eq:maximum}) coincides
with the boundaries of the region. To verify them as local maximum,
we can compare the gradient $\nabla f=\left(\begin{array}{c}
\partial_{k_{1}}f\\
\partial_{k_{2}}f\\
\partial_{l_{1}}f
\end{array}\right)=\left(\begin{array}{c}
-2(1+h+4k_{1}+2k_{2}+3l_{1}-N-2n)\\
-(1/2)-h-4k_{1}-7k_{2}-3l_{1}+N+2n\\
-3h-6k_{1}-3k_{2}-6l_{1}+N+4n
\end{array}\right)$ with the orientation of boundaries $g_{i}\leq0$. Specifically, we
solve 
\begin{equation}
\nabla(f-\sum_{i}\mu_{i}g_{i})=0\label{eq:local_maximum_condition}
\end{equation}
 and verify $\mu_{i}\ge0\quad\forall i$.

When $n<\frac{N-h}{2}$, the constraints are $\left\{ \begin{array}{c}
g_{1}=k_{1}+k_{2}+l_{1}-n\\
g_{2}=-k_{2}\\
g_{3}=-l_{1}
\end{array}\right.$. Solving Eq. (\ref{eq:local_maximum_condition}) we get 
\[
\boldsymbol{\mu}=\left(\begin{array}{c}
2(N-h-2n-1)\\
N-h-2n-(3/2)\\
N+h-2n-2
\end{array}\right)
\]

When $\frac{N-h}{2}<n<\frac{N+h}{2}$, the constraints are $\left\{ \begin{array}{c}
g_{1}=2k_{1}+k_{2}+l_{1}-(N-h)\\
g_{2}=-k_{2}\\
g_{3}=-l_{1}
\end{array}\right.$ where $g_{1}$ comes from $l_{2}\geq0$. Solving Eq. (\ref{eq:local_maximum_condition})
we get 
\[
\boldsymbol{\mu}=\left(\begin{array}{c}
2n+h-N-1\\
-1/2\\
N+h-2n-1
\end{array}\right)
\]

When $n>\frac{N+h}{2}$, the constraints are $\left\{ \begin{array}{c}
g_{1}=2k_{1}+k_{2}+l_{1}-(N-h)\\
g_{2}=-k_{2}\\
g_{3}=-k_{1}-l_{1}+n-h
\end{array}\right.$ where $g_{3}$ comes from $k_{1}+k_{2}+l_{2}\leq m$. Solving Eq.
(\ref{eq:local_maximum_condition}) we get 
\[
\boldsymbol{\mu}=\left(\begin{array}{c}
-2+h+2n-N\\
-(3/2)-h+2n-N\\
-2(1+h-2n+N)
\end{array}\right)
\]

All of $\mu_{i}$ above are non-negative.
\end{widetext}

\end{document}